\documentclass[twocolumn, aps, prb]{revtex4}
\usepackage{graphicx}
\usepackage{amssymb}
\usepackage{amsmath}
\def\Journal #1,#2,#3,#4#5#6#7{#1 {\bf #2}, #3 (#4#5#6#7)}

\def\gsim{\lower -0.3ex \hbox{$>$} \kern -0.75em \lower 0.7ex
\hbox{$\sim$}}
\def\lsim{\lower -0.3ex \hbox{$<$} \kern -0.75em \lower 0.7ex
\hbox{$\sim$}}

\def\Vec#1{{\bf #1}}

\def\vare{\varepsilon}

\begin{document}
%
\title{
Energy Spectrum and Quantum Hall Effect in Twisted Bilayer Graphene
}
\author{Pilkyung Moon and Mikito Koshino}
\affiliation{
Department of Physics, Tohoku University, 
Sendai, 980--8578, Japan}
\date{\today}

\begin{abstract}
We investigate the electronic structure
and the quantum Hall effect in twisted bilayer graphenes
with various rotation angles in the presence of magnetic field. 
Using a low-energy approximation,
which incorporates the rigorous interlayer interaction,
we computed the energy spectrum and the quantized Hall conductivity
in a wide range of magnetic field 
from the semi-classical regime to the fractal spectrum regime.
In weak magnetic fields, the low-energy conduction band 
is quantized into electronlike 
and holelike Landau levels at energies
below and above the van Hove singularity, respectively,
and the Hall conductivity sharply drops 
from positive to negative when the Fermi energy
goes through the transition point.
In increasing magnetic field, the spectrum gradually evolves into 
a fractal band structure called Hofstadter's butterfly,
where the Hall conductivity exhibits a non-monotonic behavior
as a function of Fermi energy.
The typical electron density and magnetic field amplitude
characterizing the spectrum monotonically
decrease as the rotation angle is reduced,
indicating that the rich electronic structure 
may be observed in a moderate condition. 
\end{abstract}
\maketitle
%
\section{Introduction}
\label{sec_intr}

The electronic structure of bilayer graphene is 
highly sensitive to the stacking geometry between the two layers.
The interlayer interaction in bilayer graphene with regular AB stacking
\cite{Novoselov_et_al_2005a,Zhang_et_al_2005a,Novoselov_et_al_2006a}
modifies the linear dispersion of monolayer graphene 
into the quadratic dispersion, where an electron behaves as
a massive particle.\cite{McCann_and_Falko_2006a}
On the other hand, the recent epitaxial growth
technique \cite{Berger_et_al_2006,Hass_et_al_2007a}
realized twisted bilayer graphene (TBG)
in which two layers are stacked with a random rotation angle.
\cite{Hass_et_al_2007a,Hass_et_al_2008a,Luican_et_al_2011a}
The unit cell area of TBG can be more than 1000 times 
as large as that of monolayer graphene, due to slightly
misoriented lattice vectors of two layers. Such an atomic
configuration was observed as Moir\'{e} pattern
in the scanning tunneling
microscopy.\cite{Hass_et_al_2008b,Miller_et_al_2010b,Miller_et_al_2010a,Zhao_et_al_2011a,Li_et_al_2009a}
TBG was also fabricated in different methods
such as folding of mechanically exfoliated graphenes,\cite{Ni_et_al_2008a}
segregation of graphene on Ni film,\cite{Zhao_et_al_2011a}
and unzipping of carbon nanotube.\cite{Xie_2011_1318}

The electronic structure of TBG shows
a linear band dispersion near Dirac points
 \cite{Latil_et_al_2007a,Lopes_dos_Santos_et_al_2007a,Shallcross_et_al_2008a,Hass_et_al_2008a,E_Suarez_Morell_et_al_2010a}
rather than the massive dispersion of AB stacked bilayer,
suggesting relatively weak interlayer interaction.
In strong magnetic fields, however,
it is predicted that the spectrum exhibits a fractal structure 
called Hofstadter's butterfly,
in which a series of energy gaps appears in a self-similar fashion.
\cite{Bistritzer_and_MacDonald_2011a,Hofstadter_1976a} 
The fractal band structure generally 
occurs in a periodic system when the magnetic 
flux per a unit cell is comparable to $h/e$,
and this condition is realized in TBG
in a reasonable magnetic-field range owing to the large unit cell. 
The fractal band structure and the quantum Hall effect were
theoretically studied for TBG in the strong magnetic field regime
using a continuous interlayer coupling model.
\cite{Bistritzer_and_MacDonald_2011a} 
Experimentally, the energy spectrum of the twisted graphene 
stacks in magnetic field was probed in the transport measurement
\cite{Lee_et_al_2011a,Sanchez_Yamagishi_et_al_2012a}
and the magneto-optical absorption, \cite{Crassee_et_al_2011a}
while the fractal band structure has not yet been observed.

In this paper, we investigate the electronic spectrum and the quantum
Hall effect in TBG with various rotation angles and magnetic fields.
We calculate the spectrum by including a limited number of bases
which are significant in the low-energy spectrum,
while rigorously taking account of transfer integrals
between lattice points on the different layers.
Using this method, we describe 
the spectral evolution in a wide range of magnetic field, 
from the semiclassical 
Landau levels in the weak-field regime
to the fractal band structure in the strong-field regime.

In weak magnetic fields, we find that the low-energy 
conduction band is quantized into electronlike Landau levels 
and holelike Landau levels at energies below and above the
van Hove singularity, respectively,
in accordance with the topological change of the Fermi surface
from electron-type to hole-type at the band saddle point.
As a consequence, the quantized Hall conductivity abruptly jumps
from positive to negative when the Fermi energy goes through 
the transition point.
In increasing magnetic field, the electron and hole Landau levels 
begin to be mixed and gradually evolve into the fractal band structure.
We calculate the quantized Hall conductivity for each single gap, 
and demonstrate that it changes non-monotonically 
as a function of Fermi energy and magnetic field.
\cite{Kohmoto_1985a,Thouless_et_al_1982a}

\section{THEORETICAL METHODS}
\label{sec_theo}

\subsection{Atomic structure}
\begin{figure}
\begin{center}
\leavevmode\includegraphics[width=0.7\hsize]{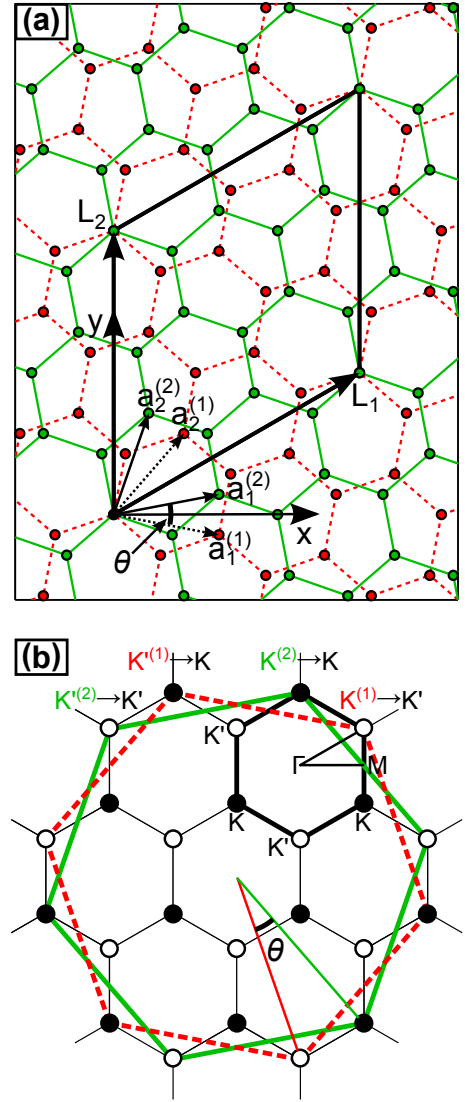}
\end{center}
\caption{ (Color online) (a) Atomic structure of TBG with rotation
angle $\theta= 21.8^\circ$. Dashed (red) and solid (green) lines represent
the lattices of layers 1 and 2, respectively. 
(b) Brillouin zone of TBG with $\theta= 21.8^\circ$.
Dashed (red) and solid (green) large hexagons 
correspond to the first Brillouin zone of layer 1 and 2, respectively, 
and thick small-hexagon to the reduced Brillouin zone of TBG. 
Open and filled circles
are two inequivalent valleys $K$ and $K'$ of TBG.}
\label{fig_atom1}
\end{figure}

TBG is characterized by the relative rotation angle $\theta$ 
and the relative translation vector between two graphene layers.
When the lattice structures of the two layers are commensurate,
we can define the primitive lattice vectors $\Vec{L}_1$ and $\Vec{L}_2$
as the least common multiples of the unit vectors on the two layers.
$\Vec{L}_1$ is written by integers $m,n,m',n'$ as
\cite{Mele_2010a}
\begin{eqnarray}
\Vec{L}_1 = m \Vec{a}_1^{(1)} + n \Vec{a}_2^{(1)}
= m' \Vec{a}_1^{(2)} + n' \Vec{a}_2^{(2)},
\end{eqnarray}
where $\Vec{a}_1^{(l)}$ and $\Vec{a}_2^{(l)}$ are the lattice vectors
of the layer $l=1,2$ defined in Fig.\ \ref{fig_atom1}(a).
$\Vec{L}_2$ is obtained by rotating $\Vec{L}_1$ by 60$^\circ$.
By appropriate choice of lattice vectors $\Vec{a}_i^{(l)}$,
the indices $(m',n')$ can be made equal to $(n,m)$, and thus
TBG is specified by a single pair of integers $(m,n)$.
The rotation angle $\theta$ is related to $(m,n)$ by
\begin{eqnarray}
 \cos\theta = \frac{1}{2}\frac{m^2+n^2+4mn}{m^2+n^2+mn},
\end{eqnarray}
and the lattice constant $L = |\Vec{L}_1| = |\Vec{L}_2|$ by 
\begin{eqnarray}
 L = a \sqrt{m^2 + n^2 + mn}
= \frac{|m-n|}{2\sin(\theta/2)} a,
\end{eqnarray}
where $a = |\Vec{a}_1| = |\Vec{a}_2| \approx 0.246\,\mathrm{nm}$ 
is the lattice constant of monolayer graphene.
The area of TBG unit cell is given by 
$S = |\Vec{L}_1 \times \Vec{L}_2| = (\sqrt{3}/2)L^2$.

Figure\ \ref{fig_atom1}(a) shows the atomic structure 
of TBG with $(m,n)=(1,2)$ and $\theta = 21.8^\circ$.
Throughout the paper, 
we set the coordinates $(x, y)$ on graphene plane
so that $y$ axis is parallel to $\Vec{L}_2$,
and $z$ to the direction perpendicular to the plane.
We ignore the relative translation between two layers,
which makes a minor difference in
the electronic structure when the unit cell is large enough. 

Figure\ \ref{fig_atom1}(b) shows the extended Brillouin zone of 
TBG with $\theta = 21.8^\circ$.
The two large hexagons represent the first Brillouin zones of 
layer 1 and 2, respectively.
$K^{(l)}$ and $K'^{(l)}$ denote the two inequivalent 
corners of layer $l$, which are Dirac points
in the single-layer band structure.
The four Dirac points of $K^{(1)}$, $K'^{(1)}$, $K^{(2)}$, and $K'^{(2)}$ are
folded back to two Dirac points, $K$ and $K'$, 
in the reduced Brillouin zone. 
\cite{Shallcross_et_al_2010a}

\begin{figure*}[ht]
\begin{center}
\leavevmode\includegraphics[width=\hsize]{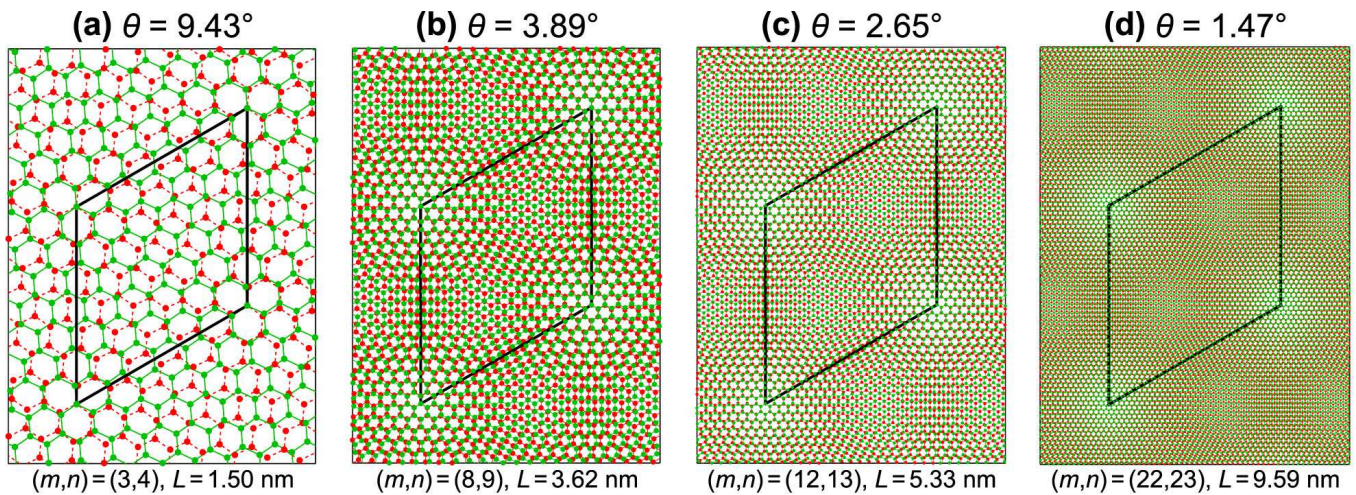}
\end{center}
\caption{ 
(Color online) Atomic structure of TBG with rotation angles of
(a) 9.43$^\circ$, (b) 3.89$^\circ$, (c) 2.65$^\circ$, and 
(d) 1.47$^\circ$. 
Dashed (red) and solid (green) lines represent
lattices of layer 1 and 2, respectively. 
$(m,n)$ is the index characterizing
the primitive lattice vector of TBG,
and $L$ is the length of the lattice vector.
}
\label{fig_atom2}
\end{figure*}

Figure\ \ref{fig_atom2} shows the atomic structures
of four different TBGs to be considered in following sections.
They are specified by $(m,n) = (3,4),(8,9), (12,13)$, and $(22,23)$,
and the rotation angles
$\theta = 9.43^\circ$, 3.89$^\circ$, 2.65$^\circ$, and
1.47$^\circ$, respectively.
As the angle $\theta$ decreases, the size of the unit cell
enlarges and the Moir\'{e} pattern becomes evident.

\subsection{Tight-binding model}

In a tight-binding model in terms of $p_z$ atomic orbitals,
the Hamiltonian of TBG at zero magnetic field is written as
\begin{eqnarray}
 H^{B=0}_{\rm TBG} = -\sum_{\langle i,j\rangle}
t(\Vec{R}_i,\Vec{R}_j)
|\Psi_i\rangle\langle\Psi_j| + {\rm H.c.},
\label{eq_H_TBG}
\end{eqnarray}
where $\Vec{R}_i$ and $|\Psi_i\rangle$ 
represent the lattice point and the atomic state at site $i$, respectively,
and $t(\Vec{R}_i,\Vec{R}_j)$ is
the transfer integral between the sites $i$ and $j$. 
We adopt an approximation,
\cite{Nakanishi_and_Ando_2001a,Uryu_2004a,Trambly_de_Laissardiere_et_al_2010a,Slater_and_Koster_1954a}
\begin{eqnarray}
 && -t(\Vec{R}_i,\Vec{R}_j) = 
V_{pp\pi}\left[1-\left(\frac{\Vec{d}\cdot\Vec{e}_z}{d}\right)^2\right]
+ V_{pp\sigma}\left(\frac{\Vec{d}\cdot\Vec{e}_z}{d}\right)^2,
\nonumber \\
&& V_{pp\pi} =  V_{pp\pi}^0 
\exp \left(- \frac{d-a_0}{\delta}\right),
\nonumber \\
&& V_{pp\sigma} =  V_{pp\sigma}^0 
 \exp \left(- \frac{d-d_0}{\delta}\right),
\end{eqnarray}
where $\Vec{d} = \Vec{R}_i-\Vec{R}_j$, 
and $\Vec{e}_z$ is the unit vector parallel to $z$ axis.
$V_{pp\pi}^0$ is the transfer integral between the
the nearest-neighbor atoms of monolayer graphene
which are located at distance $a_0 = a/\sqrt{3} \approx 0.142\,\mathrm{nm}$,
and $V_{pp\sigma}^0$ is the interlayer transfer integral
between vertically located atoms at the interlayer distance
$d_0 \approx 0.335\,\mathrm{nm}$.
Here we take $V_{pp\pi}^0 \approx -2.7\,\mathrm{eV}$,
$ V_{pp\sigma}^0 \approx 0.48\,\mathrm{eV}$, to
fit the low-energy dispersion of bulk graphite.
$\delta$ is the decay length of the transfer integral,
and is chosen as $0.184 a$ so that 
the next nearest intralayer coupling becomes $0.1 V_{pp\pi}^0$.
\cite{Uryu_2004a,Trambly_de_Laissardiere_et_al_2010a} 
The transfer integral for $d > 4 a_0$ is exponentially small 
and can be safely neglected.
The band velocity of the Dirac cone in monolayer graphene is given by
\begin{eqnarray}
 v \approx \frac{\sqrt{3}}{2}\frac{V_{pp\pi}^0}{\hbar}.
\end{eqnarray}

\begin{figure*}[ht]
\begin{center}
\leavevmode\includegraphics[width=\hsize]{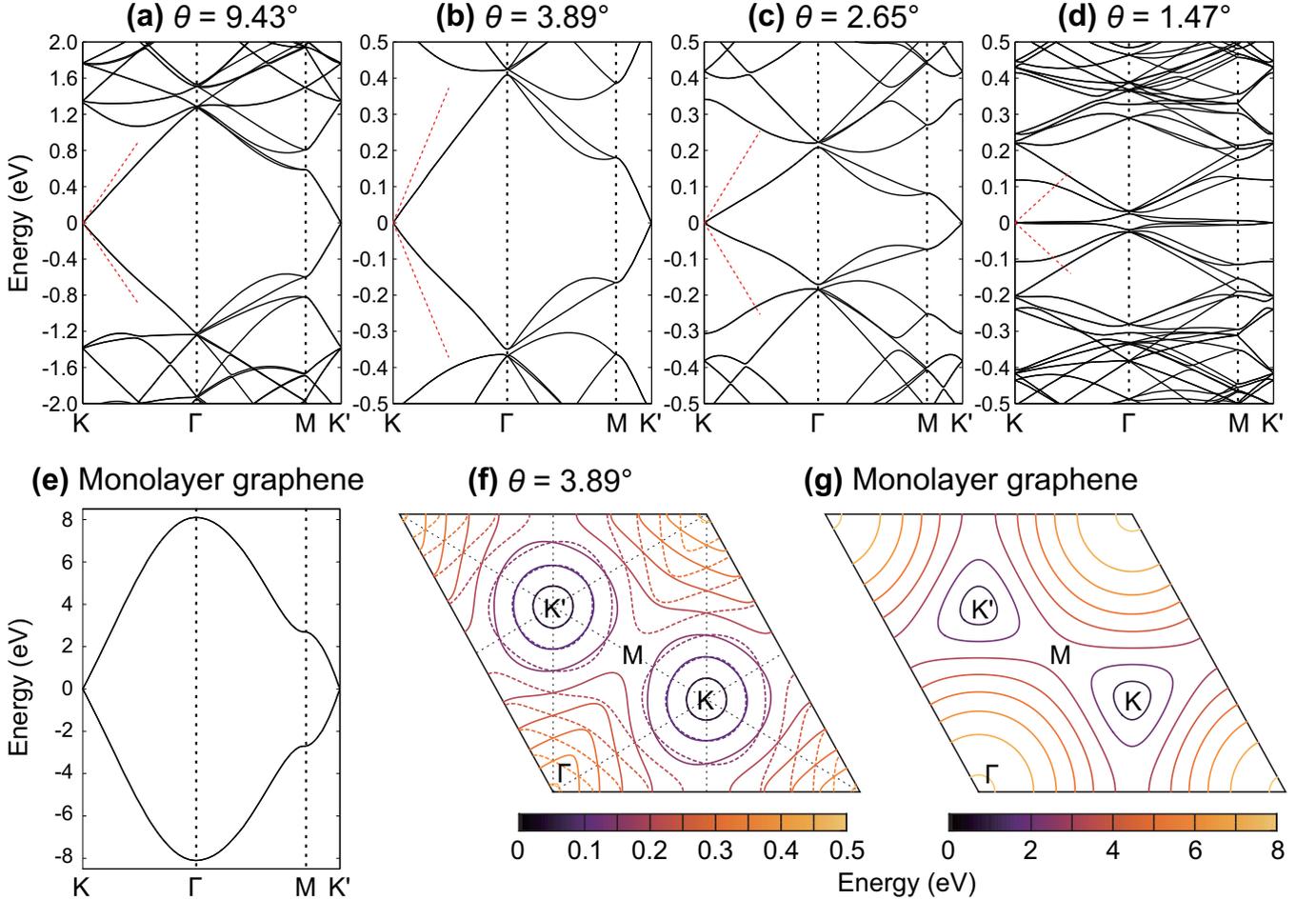}
\end{center}
\caption{
Band structure of TBG with rotation angles
(a) 9.43$^\circ$, (b) 3.89$^\circ$, (c) 2.65$^\circ$,
(d) 1.47$^\circ$, and that of (e) monolayer graphene.
Dashed (red) slopes around $K$ point
indicate the monolayer's band dispersion.
Note that the scale of wave number (horizontal axis)
reduces as the rotation angle decreases.
Dirac point energy is set to zero.
(f) Contour plot of the two lowest conduction bands
of TBG with $\theta = 3.89^\circ$.
(g) Corresponding plot for the conduction band of monolayer graphene.
}
\label{fig_band}
\end{figure*}

We plot the energy bands of four TBGs with the
different rotation angles in Figs.\ \ref{fig_band}(a)-(d).
Dashed (red) lines near $K$ point 
indicate the band dispersion of monolayer graphene,
of which the entire structure is shown in Fig.\ \ref{fig_band}(e).
The low-energy spectrum can be understood by  
folding monolayer's Dirac cone into the reduced Brillouin zone,
and thus the structures are similar among different rotation angles
except for the scale.
The lowest band is characterized by
a linear dispersion analogous to monolayer graphene
at the $K$ and $K'$ points,\cite{Latil_et_al_2007a,Lopes_dos_Santos_et_al_2007a,Shallcross_et_al_2008a,Hass_et_al_2008a,E_Suarez_Morell_et_al_2010a}
the van Hove singularity 
at the $M$ point,\cite{Ni_et_al_2009a,Li_et_al_2009a,E_Suarez_Morell_et_al_2010a,Wang_et_al_2010a}
and a holelike pocket at the $\Gamma$ point.
In accordance with the band folding picture, the width of 
the lowest band is roughly given by $4\pi \hbar v / (3L)$, which is
the graphene's band gradient times the distance between $K$ and
$\Gamma$.
In small rotation angles less than 5$^\circ$, however, 
the width becomes significantly smaller than this estimate
because the level repulsion from the upper bands
becomes comparable to the band width itself.
As a result, the velocity of the Dirac cone gradually reduces
from the monolayer's $v$.
In the smallest rotation angle $\theta = 1.47^\circ$,
in particular, the lowest energy band is highly distorted,
and nearly flat dispersion appears near zero energy.
\cite{Trambly_de_Laissardiere_et_al_2010a,Bistritzer_and_MacDonald_2011b}

The lowest band of TBG is composed of 
a pair of nearly degenerate branches. 
Figure \ref{fig_band}(f) shows the contour plots of the 
two lowest conduction bands in  $\theta = 3.89^\circ$. 
Those two bands, indicated by solid and broken curves, 
are mirror symmetric to each other with respect to the lines of
$K-\Gamma$, $K'-\Gamma$, and $K-K'$,
reflecting the $C_2$ symmetry in the real-space
lattice structure. 
Each of the two bands has a similar landscape to that of monolayer graphene
which is shown in Fig.\ \ref{fig_band} (g),
where the linear dispersion, the saddle point and
the hole pocket appear near $K(K')$, $M$, and $\Gamma$, respectively.

\subsection{Electronic structures in magnetic fields}

We consider TBG in a uniform magnetic field
$\Vec{B} = (0,0,B)$ perpendicular to the layer.
For simplicity, we neglect spin Zeeman splitting throughout the
paper. The system is characterized by 
the number of magnetic flux per a unit cell,
$\Phi = BS$,
measured in units of the flux quantum $\Phi_0 = h/e$.
In the magnetic field, the Hamiltonian is no longer translationally-symmetric
because of the spatial dependence of the vector potential.
When $\Phi/\Phi_0$ is a rational number $p/q$ ($p$ and $q$ 
are coprime integers), however, 
we can introduce a magnetic
unit cell with lattice vectors $\tilde{\Vec{L}}_1 = q \Vec{L}_1$
and $\tilde{\Vec{L}}_2 = \Vec{L}_2$, and construct the eigenstates
so as to satisfy the magnetic Bloch condition.\cite{Brown_1968a,Xial_et_al_2010a}
By choosing the vector potential 
as $\Vec{A} = (0,Bx,0)$ and
taking the $y$ axis parallel to $\Vec{L}_2$, 
the magnetic Bloch condition for TBG is written as
\begin{eqnarray}
 \Psi_\Vec{k}(\Vec{r}+\tilde{\Vec{L}}_1) &=& 
e^{i \Vec{k}\cdot\tilde{\Vec{L}}_1}
e^{-i (e/\hbar)(\Vec{A}- \Vec{B}\times \Vec{r})\cdot\tilde{\Vec{L}}_1}
 \Psi_\Vec{k}(\Vec{r}), 
\nonumber\\
 \Psi_\Vec{k}(\Vec{r}+\tilde{\Vec{L}}_2) &=& 
e^{i \Vec{k}\cdot\tilde{\Vec{L}}_2}
 \Psi_\Vec{k}(\Vec{r}),
\label{eq_mag_Bloch}
\end{eqnarray}
where $\Vec{k}$ is the Bloch wavenumber
defined in the magnetic Brillouin zone spanned by reciprocal vectors of
$\tilde{\Vec{L}}_1$ and $\tilde{\Vec{L}}_2$.
Since the magnetic unit cell is $q$ times as large as the unit
cell in the absence of magnetic field, the magnetic Brillouin zone is
$q$-fold of the original, and each energy band at zero magnetic field
splits into $q$ subbands.\cite{Hofstadter_1976a}

The tight-binding Hamiltonian under a magnetic field is obtained by
adding a phase factor to the transfer integral 
in Eq.\ (\ref{eq_H_TBG}). This is written as
\begin{eqnarray}
&& H_{\rm TBG} = -\sum_{\langle i,j\rangle}
t(\Vec{R}_i,\Vec{R}_j) e^{i\phi_{ij}}
|\Psi_i\rangle\langle\Psi_j| + {\rm H.c.},\nonumber\\
&&
\phi_{ij} = -\frac{e}{\hbar}\int_\Vec{R_j}
^\Vec{R_i}\Vec{A}(\Vec{r})\cdot d\Vec{r}.
\end{eqnarray}
It is, however, not practical to calculate
the energy spectrum of TBG by diagonalizing this Hamiltonian, 
since the number of atoms in a magnetic unit cell 
is huge in feasible magnetic fields. 
Instead, we construct the basis
from the effective mass wavefunctions for Landau levels
of monolayer graphene,
which approximate the eigenstates in the absence of
the interlayer coupling.
We then truncate the bases far from the Dirac point,
and compose the Hamiltonian matrix by writing 
$H_{\rm TBG}$ in terms of the reduced basis.

In monolayer graphene under magnetic field,
the eigenstates are labeled by $(v,n,k_y)$
with the valley index $v=K,K'$, the Landau level index
$n = 0,\pm 1, ...$, and the wave vector $k_y$ along
$y$ direction.
\cite{McClure_1956a,Shon_and_Ando_1998a,Zheng_and_Ando_2002a,
Gusynin_and_Sharapov_2005a,Ando_2005a}
The eigenenergy depends only on $n$ as
\begin{eqnarray}
\varepsilon_n = \hbar\omega_B\,\,{\rm sgn}(n)\sqrt{|n|},
\end{eqnarray}
with $\hbar\omega_B = \sqrt{2\hbar v^2 e B}$.
The effective wavefunctions are written as
\cite{Shon_and_Ando_1998a,Zheng_and_Ando_2002a}
\begin{eqnarray}
 \Vec{F}_{Knk_y}(\Vec{r}) &=& \frac{C_n}{\sqrt{L}}e^{ik_y y}
\left(
\begin{array}{c}
{\rm sgn}(n) (-i) \phi_{|n|-1,k_y}(x)\\
\phi_{|n|,k_y}(x)\\0\\0
\end{array}
\right), \nonumber\\
 \Vec{F}_{K'n k_y}(\Vec{r}) &=& \frac{C_n}{\sqrt{L}}e^{ik_y y}
\left(
\begin{array}{c}
0\\0\\
\phi_{|n|,k_y}(x)\\
{\rm sgn}(n) (-i) \phi_{|n|-1,k_y}(x)
\end{array}
\right).
\label{eq_mono_LL}
\end{eqnarray}
Here $\Vec{F} = (F^K_A,F^K_B,F^{K'}_A,F^{K'}_B)$ 
is a four-component vector representing 
the envelope function of each site and valley.
We defined
$\phi_{n,k}(x) = (2^{n}n!\sqrt{\pi}l_B)^{-1/2} \,\,e^{-z^2/2}H_{n}(z)$,
with $z = (x+kl_B^2)/l_B$ and $H_n$ being the Hermite
polynomial, $l_B = \sqrt{\hbar/(eB)}$, and
\begin{eqnarray}
 C_n = \left\{
\begin{array}{cc}
 1 & (n=0), \\
 1/\sqrt{2} & (n\neq 0),
\end{array}
\right.
\nonumber\\
 {\rm sgn}(n) = \left\{
\begin{array}{cc}
 0 & (n=0), \\
 n/|n| & (n\neq 0).
\end{array}
\right.
\end{eqnarray}

The tight-binding wavefunction $\Psi$ on the layer $l$ can be expressed
in terms of the envelope function $\Vec{F}$ as \cite{Ando_2005a}
\begin{eqnarray}
\Psi_A(\Vec{R}_A) &=&
e^{i\Vec{K}^{(l)}\cdot\Vec{R}_A}F^{K}_A (\Vec{R}_A)
+ e^{i\eta^{(l)}} e^{i\Vec{K}'^{(l)}\cdot\Vec{R}_A}F^{K'}_A (\Vec{R}_A)
\nonumber\\
\Psi_B(\Vec{R}_B) &=&
-\omega e^{i\eta^{(l)}} e^{i\Vec{K}^{(l)}\cdot\Vec{R}_B}F^{K}_B (\Vec{R}_B)
\nonumber\\
&& \hspace{25mm} +  e^{i\Vec{K}'^{(l)}\cdot\Vec{R}_B}F^{K'}_B (\Vec{R}_B),
\end{eqnarray}
where $\eta^{(l)}$ is the angle of
$\Vec{a}_1^{(l)}$ to $x$ axis.
We define $\Psi^{(l)}_{v n k_y}$ as the tight-binding wavefunction 
on the layer $l$ generated from $\Vec{F}_{v n k_y}$.

We then combine the bases of different $k_y$ 
so as to satisfy the magnetic Bloch condition, Eq.\ (\ref{eq_mag_Bloch}).
We define
\begin{eqnarray}
\Psi^{(l)}_{v n \Vec{k} m} &=& 
\sum_{j=-\infty}^\infty 
\alpha^j
\exp\left[i\pi p q \frac{j(j+1)}{2}\right]
\Psi^{(l)}_{v n k_y^{(m)}},
\nonumber\\
\alpha &=& 
\exp\left[i(\Vec{k}-\Vec{K}^{(l)}_v)
\cdot\left(\tilde{\Vec{L}}_1-\frac{q}{2}\tilde{\Vec{L}}_2\right)\right]
\nonumber \\
k_y^{(m)} &=& k_y - (\Vec{K}^{(l)}_v)_y - \frac{2\pi}{L_y}(pj+m),
\end{eqnarray}
where $\Vec{k}$ is the Bloch wave number
defined in the magnetic Brillouin zone,
$m=0,1,\cdots p-1$,
and $\Vec{K}^{(l)}_v$ represent $\Vec{K}^{(l)}, \Vec{K}'^{(l)}$
for $v=K,K'$, respectively.
It is straightforward to show that this satisfies 
the condition of Eq.\ (\ref{eq_mag_Bloch}).

An eigenstate of TBG is written as
a linear combination of single-layer eigenstates
$\Psi^{(l)}_{v n \Vec{k} m}$ belonging to the same $\Vec{k}$.
We only include single-layer bases within 
$-E_{\rm max} < \vare_n < E_{\rm max}$,
to discard the bases which do not much affect the low-energy spectrum.
The eigenenergies are obtained by diagonalizing the Hamiltonian matrix 
within the reduced bases,
\begin{eqnarray}
 H_\Vec{k}\left[(l,v,n,m),(l',v',n',m')\right]
\equiv
\langle \Psi^{(l)}_{v n \Vec{k} m} |
H_{\rm TBG}
| \Psi^{(l')}_{v' n' \Vec{k} m'} \rangle,
\nonumber\\
\label{eq_H_reduced}
\end{eqnarray}
for each $\Vec{k}$ in the magnetic Brillouin zone.
The cut-off energy should be sufficiently larger than the
interlayer-coupling energy, which is of the order of $V_{pp\sigma}^0$
at most, and tends to decrease in small twisting angles.
Here we take $E_{\rm max}= 1.5\,\mathrm{eV}$ for $\theta = 9.43^\circ$ 
and 3.89$^\circ$, and 1.0\,eV for 2.65$^\circ$ and 1.47$^\circ$.
To avoid undesired effects 
caused by a discrete change in the number of bases
in varying magnetic field, we adopt a soft cut-off 
which gradually reduces the matrix elements
associated to the single-layer bases beyond $\pm E_{\rm max}$.

We calculate the matrix elements [Eq.\ (\ref{eq_H_reduced})] 
between different layers by evaluating the transfer integral
for each pair of carbon atoms 
up to the cut-off distance $d = 4a_0$. 
The matrix elements within the same layer
can be replaced with a diagonal matrix composed
of the effective-mass eigenenergies in monolayer graphene,
\begin{eqnarray}
  H_\Vec{k}\left[(l,v,n,m),(l,v',n',m')\right]
= \vare_n \, \delta_{v,v'}\delta_{n,n'}\delta_{m,m'}.
\end{eqnarray}
This treatment is valid in low energies,
as long as the magnetic field is not too strong, or $l_B \gg a$.

When the Fermi energy $\vare_F$ is inside a band gap of the spectrum,
the Hall conductivity $\sigma_{xy}$
is evaluated by the formula\cite{Streda_1982a,Widom_1982a}
\begin{eqnarray}
 \sigma_{xy} = -e\left(\frac{\partial n_F}{\partial B}\right)_{\vare_F},
\end{eqnarray}	 
where $n_F$ is the electron density per unit area below the gap.

\begin{figure*}
\begin{center}
\leavevmode\includegraphics[width=0.9\hsize]{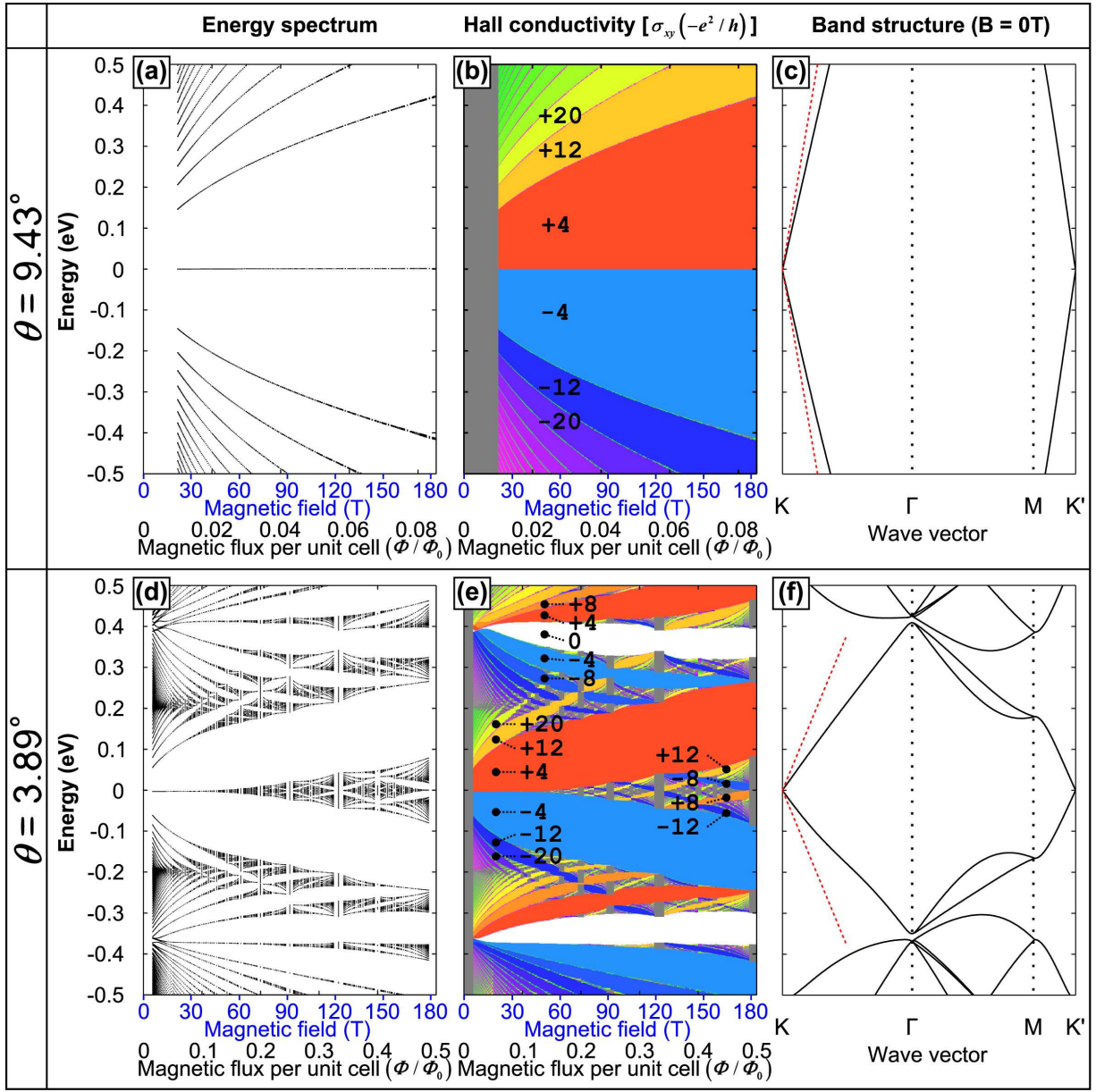}
\end{center}
\caption{ 
(Color online)
Energy spectrum and quantum Hall effect 
in TBG in magnetic field
with rotation angles of 9.43$^\circ$ (above) and 3.89$^\circ$ (below).
In each row, the left and middle panels display
the energy spectrum and the quantized Hall conductivity 
as functions of magnetic field strength, respectively,
and the right panel shows the band structure 
in the absence of magnetic field.
Dashed (red) slopes around the $K$ point
indicate the dispersion of monolayer graphene.
In (b) and (e), the quantized values of Hall conductivity 
inside energy gaps are indicated by numbers as well as colors
filling the gaps.
The Hall conductivity of the gray area 
cannot be determined by the present calculation.
}
\label{fig_spec1}
\end{figure*}

\begin{figure*}
\begin{center}
\leavevmode\includegraphics[width=0.9\hsize]{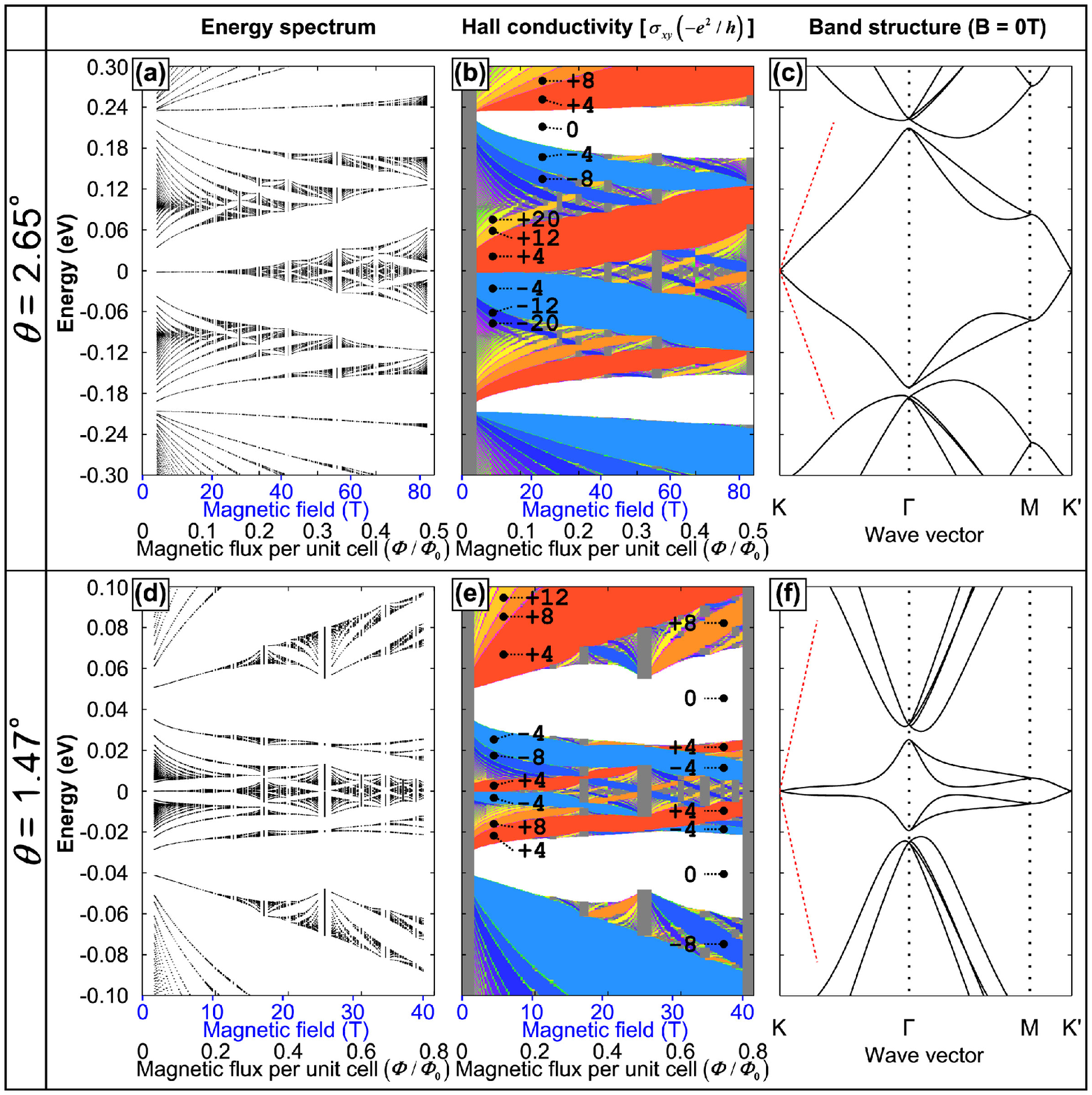}
\end{center}
\caption{ 
(Color online) Plots similar to Fig.\ \ref{fig_spec1}
for TBG with rotation angles of 
2.65$^\circ$ (above) and 1.47$^\circ$ (below).
}
\label{fig_spec2}
\end{figure*}

\section{RESULTS AND DISCUSSION}
\label{sec_results}

We show the energy spectrum (left) and 
quantized Hall conductivity (middle) 
against the magnetic field amplitude,
for $\theta= 9.43^\circ$, 3.89$^\circ$
in Fig. \ref{fig_spec1},
and for 2.65$^\circ$, 1.47$^\circ$ in Fig. \ref{fig_spec2}.
In the right-most panel, we show the zero-field band structure
in the same energy range.
The energy spectrum of $\theta = 9.43^\circ$ [Fig.\ \ref{fig_spec1}(a)]
is almost equivalent to monolayer's Landau level,
suggesting that two layers are nearly decoupled in this energy region.
The sequence of the Hall conductivity, 
$4, 12, 20, \cdots$ in units of $-e^2/h$,
\cite{de_Gail_et_al_2011a,Choi_et_al_2011a,Lee_et_al_2011a} 
is exactly twice as large as the monolayer's.
\cite{Zheng_and_Ando_2002a,Novoselov_et_al_2005a,Zhang_et_al_2005a,Gusynin_and_Sharapov_2005a}
Each Landau level is eight-fold degenerate
due to the number of layers as well as 
the spin and valley degeneracies.

In contrast, the energy spectrum of $\theta = 3.89^\circ$ 
[Figs.\ \ref{fig_spec1}(d) and \ref{fig_spec1}(e)] 
exhibits a complicated structure which is clearly
distinguished from monolayer graphene.
In weak magnetic fields of $\Phi/\Phi_0 < 0.1$,
the low-energy spectrum below 0.2\,eV
shows monolayerlike Landau levels
and Hall conductivity of $4, 12, 20, \cdots$.
In the higher energy region above 0.2\,eV, 
on the other hand, we observe holelike Landau levels moving
downward in energy, and the negative Hall conductivity of
$0, -4, -8, -12, \cdots$.
When the electron density increases from the charge neutrality point, 
the Hall conductivity rises in a sequence of $4, 12, 20, \cdots$
with a step of 8,
then abruptly drops to a negative
extremum, and increases with a step of 4 all the way to zero.

Those spectral features in weak magnetic field
perfectly coincide with the zero-field band structure
in Fig.\ \ref{fig_spec1}(f).
The electronlike Landau levels 
are regarded as the quantized orbits accommodated
in electron pockets at $K$ and $K'$ points,
while the holelike Landau levels are those 
in a hole pocket at $\Gamma$ point.
The transition from electronlike levels to holelike levels 
corresponds to topological change of the Fermi surface 
at the saddle point ($M$ point),
which is responsible for the van Hove singularity at 0.2\,eV.
The step of the Hall conductivity 
reflects the number of electron and hole pockets in the 
first Brillouin zone, i.e., 
the degeneracy of an electronlike level
is twice as large as that of a holelike level,
because there are inequivalent $K$ and $K'$ points
whereas there is only one $\Gamma$ point.
Note that the pair of nearly degenerate 
lowest conduction bands [Fig.\ \ref{fig_band}(f)]
give the identical Landau level energies
and contribute to the degeneracy of two 
in addition to the spin degeneracy.
Except for this doubling,
the low-energy Landau level spectrum and the quantized Hall conductivity 
of TBG are quite analogous to 
those of whole $\pi$ band in monolayer graphene, \cite{Hatsugai_et_al_2006a}
as expected from the similarity 
of the band structure at zero magnetic field.

\begin{figure}
\begin{center}
\leavevmode\includegraphics[width=0.95\hsize]{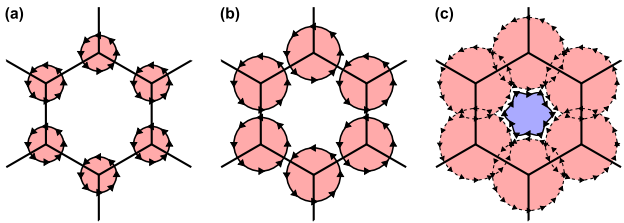}
\end{center}
\caption{ 
(Color online) Fermi circle and electron trajectories 
of TBG in a nearly free electron picture,
for three different Fermi energies
(a) in the vicinity of Dirac points, 
(b) near van Hove singularity at the saddle point, and (c) holelike
band at the $\Gamma$ point.
}
\label{fig_FS}
\end{figure}

The electronlike and holelike Landau levels
are alternatively explained by a nearly-free electron model,
without mentioning the rigorous zero-field band structure. 
In Fig.\ \ref{fig_FS}, we illustrate 
semi-classical electron trajectories 
at several different Fermi energies
for a ``free'' TBG with interlayer coupling neglected.
In the limit of a small Fermi energy [Fig.\ \ref{fig_FS}(a)], 
electrons move along closed orbits around $K$ and $K'$,
and those motions are quantized into monolayerlike Landau levels.
Since each of $K$ and $K'$ points include two original $K$ points
from top and bottom monolayers, the Hall conductivity
yields $4, 12, 20, \cdots$,
i.e., double of monolayer's sequence.
For large Fermi energies, the electron orbits 
around the $K$ and $K'$ valleys cross each other
as shown in Fig.\ \ref{fig_FS}(c).
A finite interlayer coupling interchanges the orbits at each 
crossing point,
and generates a single holelike trajectory 
moving around $\Gamma$ point in the opposite direction. 
The corresponding holelike Landau levels are
four-fold degenerate due to spin and the Fermi circle doubling,
and thus the Hall conductivity takes $0, -4, -8, -12, \cdots$.
The middle panel [Fig.\ \ref{fig_FS}(b)] is for
the intermediate energy region between two regimes.
There, the different semiclassical orbits are strongly mixed
by the magnetic breakdown due to a small $k$-space separation,
resulting in broadening of Landau levels near
the van Hove singularity in Fig.\ \ref{fig_spec1}(d).

The electron density to fill the lowest conduction band is given by
\begin{eqnarray}
 n_0 = \frac{2 g_s}{S},
\end{eqnarray}
where $g_s$ is the spin degeneracy and 2 is the band doubling.
$n_0$ characterizes the order of the electron density 
required to reach
the van Hove singularity and the holelike Landau levels.
We have $n_0 = 3.5, 1.6$ and 0.5 in units of $10^{13}$\,cm$^{-2}$
for $\theta = 3.89^\circ, 2.65^\circ$ and 
$1.47^\circ$, respectively.
In monolayer graphene,
the electron density to access the van Hove singularity
is of the order of $10^{15}$\,cm$^{-2}$.

The semiclassical picture breaks down
when the magnetic field is so strong that
\begin{eqnarray}
 l_B \lsim L,
\label{eq_cond}
\end{eqnarray}
because then the uncertainty in electron momentum
($\sim 2\pi/l_B$) becomes comparable or larger
than the size of the Brillouin zone ($\sim 2\pi/L$),
and a semiclassical cyclotron orbit is not well defined
anymore.
Then the energy spectrum, including even $n = 0$ Landau level, exhibits 
a fractal band structure.
\cite{Bistritzer_and_MacDonald_2011a, Hofstadter_1976a}
The magnetic field strength needed to observe a fractal 
structure becomes more feasible in smaller rotation angles,
due to larger unit cell size $L$.
The condition $l_B \lsim L$ is equivalent to 
$\Phi/\Phi_0 \gsim \sqrt{3}/(4\pi) \approx 0.14$,
which amounts to $B \gsim 50$\,T, 23\,T and 7.2\,T 
for $\theta = 3.89^\circ, 2.65^\circ$ and $1.47^\circ$, 
respectively.
In Fig.\ \ref{fig_spec1}, we actually observe that the electron and hole 
Landau levels gradually evolve into the fractal structure
as the magnetic field exceeds the critical value.
The Hall conductivity in the fractal regime
behaves non-monotonically as a function of Fermi energy.
\cite{Kohmoto_1985a,Thouless_et_al_1982a}

The energy spectrum of $\theta = 3.89^\circ$
and that of $\theta = 2.65^\circ$
(Fig.\ \ref{fig_spec2}) exhibit similar structures
except for the energy scale,
as expected from the resemblance between the band structures
argued in the previous section.
In the case of $\theta=1.47^\circ$, the spectrum
is strongly compressed in the vicinity of Dirac
points, in accordance with
the band width reduction in small rotation angles.
Although the band structure near Dirac points is almost flat,
$\Gamma$ point still has a finite band velocity which is about $0.6v$.
As a consequence, the energy gaps between the 
holelike Landau levels are much wider than those between
the electronlike levels.

While we have considered some specific commensurate angles,
a similar fractal energy spectrum should
appear 
in any small angles including incommensurate ones, 
as long as the lattice structure exhibits 
a long-period Moir\'{e} pattern.
As a natural extension of the previous argument,
the condition for the fractal spectrum in general angles
is expected to be
\begin{eqnarray}
 l_B \lsim L_{\rm M}
\label{eq_cond_incommensurate}
\end{eqnarray}
instead of Eq.\ (\ref{eq_cond}),
where $L_{\rm M}$ is
the period of the Moir\'{e} pattern given by
\cite{Shallcross_et_al_2010a, Green_and_Weigle_1948}
\begin{eqnarray}
L_{\rm M} = \frac{a}{2 \sin(\theta/2)}.
\end{eqnarray}
Note that $L_{\rm M}$ is a continuous function of $\theta$,
while the rigorous unit cell size $L$
discontinuously changes depending on 
the commensurability of lattice periods,
and diverges in incommensurate angles.
$L_{\rm M}$ coincides with $L$
only in commensurate angles with $|m-n|=1$,
which are the cases considered in this paper.
The condition of Eq.\ (\ref{eq_cond_incommensurate}) 
is rewritten as
\begin{eqnarray}
 B & \gsim & \frac{4 \hbar}{e a^2} \sin^2 \frac{\theta}{2}
\approx 3.3(\textrm{T}) \times [\theta (\textrm{degree})]^2,
\end{eqnarray}
which quantifies the magnetic field required for the fractal spectrum
as a function of the rotation angle.

\section{CONCLUSION}
\label{sec_concl}

We investigated the electronic structure
and the quantum Hall effect in TBG with various rotation angles
in the presence of magnetic field.
We calculated the energy spectrum and quantized Hall conductivity
in a wide magnetic-field range, and
described the evolution 
from the semi-classical Landau levels to the fractal band structure.
In weak magnetic field, the low-energy conduction band
is quantized into electronlike and holelike Landau levels 
in accordance with the structure of the folded energy band. 
In increasing magnetic field, 
those semiclassical levels gradually evolve into Hofstadter's butterfly, 
where the Hall conductivity exhibits a non-monotonic behavior
as a function of Fermi energy.
The typical electron density and magnetic field amplitude
characterizing the spectrum monotonically
decrease as the rotation angle is reduced,
indicating that
the rich electronic properties may be observed in a moderate condition
for TBG with small angle less than $5^\circ$.

\bigskip

\section*{ACKNOWLEDGMENTS}
This work was supported by JST-EPSRC Japan-UK Cooperative
Programme Grant No. EP/H025804/1.
P.\ M.\ acknowledges the support from 
Grant-in-Aid for Research Activity Start-up (23840004) by
Japan Society for the Promotion of Science (JSPS),
and appreciates the support from Korea Institute of Science and
Technology Information Supercomputing Center through the
strategic support program for the supercomputing application
research (Grant No. KSC-2009-S02-0009),
and the Supercomputer Center, Institute for Solid State Physics,
University of Tokyo for the use of the facilities (Project No. ID: H23-D-0009).

\end{document}